\documentclass[a4paper,12pt]{article}
\usepackage{graphicx}
\def\sigmah{\sigma_h^0}
\def\Del{\Delta}

\def\ddelg{\Del\ov{\delta}_G^{}}

\def\dgbl{\Del g_L^{b}}

\def\alps{\alpha_s}

\def\zblbl{Z b_L^{} b_L^{}}

\def\zff{Zff}

\def\simgt{\,{\rlap{\lower 3.5pt\hbox{$\mathchar\sim$}}\raise 1pt\hbox{$>$}}\,}
\def\simlt{\,{\rlap{\lower 3.5pt\hbox{$\mathchar\sim$}}\raise 1pt\hbox{$<$}}\,}

\def\bite{\begin{itemize}}
\def\eite{\end{itemize}}

\def\wt{\widetilde}

\def\vsk#1{\noalign{\vskip#1 cm}}

\def\ov{\overline}
\def\disp{\displaystyle}
\def\smr{{\rm SM}}
\def\hph{\hphantom{-}}

\def\gev{~{\rm GeV}}
\def\tev{~{\rm TeV}}

\def\mt{m_t^{}}

\def\mh{m_{H_{\smr}}}
\def\xa{x_\alpha^{}}

\def\mw{m_W^{}}

\def\mz{m_Z^{}}
\def\mzsq{m_Z^2}

\def\dsz{\Del S_Z}
\def\dtz{\Del T_Z}

\def\dr{\Del R}
\def\sz{S_Z}
\def\tz{T_Z}

\def\ds{\Del S}
\def\dt{\Del T}
\def\du{\Del U}
\def\dgbl{\Del g_L^b}
\def\dmw{\Del m_W^{}}

\def\etal{{\it et al.~}}
\def\ie{{\it i.e.}}

\newcommand{\beq}{\begin{equation}}
\newcommand{\eeq}{\end{equation}}
\newcommand{\bea}{\begin{eqnarray}}
\newcommand{\eea}{\end{eqnarray}}
\newcommand{\bsub}{\begin{subequations}}
\newcommand{\esub}{\end{subequations} \noindent}

\def\PRD#1#2#3{Phys. Rev. {\bf D#1} (#2) #3}

\def\NPB#1#2#3{Nucl. Phys. {\bf B#1} (#2) #3}

\def\ZPC#1#2#3{Z. Phys. {\bf C#1} (#2) #3}
\def\EPJC#1#2#3{Eur. Phys. J. {\bf C#1} (#2) #3}
\def\PLB#1#2#3{Phys. Lett. {\bf B#1} (#2) #3}
\def\PRL#1#2#3{Phys. Rev. Lett. {\bf #1} (#2) #3}

\makeatletter
\newtoks\@stequation

\def\subequations{\refstepcounter{equation}%
  \edef\@savedequation{\the\c@equation}%
  \@stequation=\expandafter{\theequation}%   %only want \theequation
  \edef\@savedtheequation{\the\@stequation}% %expanded once
  \edef\oldtheequation{\theequation}%
  \setcounter{equation}{0}%
  \def\theequation{\oldtheequation\alph{equation}}}

\def\endsubequations{%
  \ifnum\c@equation < 2 \@warning{Only \the\c@equation\space subequation
    used in equation \@savedequation}\fi
  \setcounter{equation}{\@savedequation}%
  \@stequation=\expandafter{\@savedtheequation}%
  \edef\theequation{\the\@stequation}%
  \global\@ignoretrue}

\def\eqnarray{\stepcounter{equation}\let\@currentlabel\theequation
\global\@eqnswtrue\m@th
\global\@eqcnt\z@\tabskip\@centering\let\\\@eqncr
$$\halign to\displaywidth\bgroup\@eqnsel\hskip\@centering
     $\displaystyle\tabskip\z@{##}$&\global\@eqcnt\@ne
      \hfil$\;{##}\;$\hfil
     &\global\@eqcnt\tw@ $\displaystyle\tabskip\z@{##}$\hfil
   \tabskip\@centering&\llap{##}\tabskip\z@\cr}

\makeatother

\begin{document}
\thispagestyle{empty}
\vspace*{-15mm}
%----------
\baselineskip 10pt
\begin{flushright}
\begin{tabular}{l}
{\bf OCHA-PP-176}\\
{\bf KEK-TH-765}\\
{\bf hep-ph/0105037}
\end{tabular}
\end{flushright}
\baselineskip 24pt 
\vglue 10mm 
%%%%%%%%%%%%%%%%%%%%%%%%%%%%%%%%%%%%%%%%%%%%%%%%%%%%%%%%%%
%%%%%%%%%       TITLE      %%%%%%%%%%%%%%%%%%%%%%%%%%%%%%%
%%%%%%%%%%%%%%%%%%%%%%%%%%%%%%%%%%%%%%%%%%%%%%%%%%%%%%%%%%
\begin{center}
{\Large\bf
Supersymmetric contributions to muon $g-2$ and 
the electroweak precision measurements} 
\\
\vspace{8mm}

\baselineskip 18pt 
\def\thefootnote{\fnsymbol{footnote}}
\setcounter{footnote}{0}
{\bf Gi-Chol Cho$^{1)}$ and Kaoru Hagiwara$^{2)}$ 
\vspace{5mm}

$^{1)}${\it 
Department of Physics, Ochanomizu University, Tokyo 112-8610, Japan}\\
$^{2)}${\it Theory Group, KEK, Tsukuba, Ibaraki 305-0801, Japan}\\
}
\vspace{10mm}
\end{center}
%%%%%%%%%%%%%%%%%%%%%%%%%%%%%%%%%%%%%%%%
%%%%%                              %%%%%
%%%%%          Abstract            %%%%%
%%%%%                              %%%%%
%%%%%%%%%%%%%%%%%%%%%%%%%%%%%%%%%%%%%%%%
\begin{center}
{\bf Abstract}\\[7mm]
\begin{minipage}{14cm}
\baselineskip 16pt
\noindent
%%%%%----------------------------------
In view of the recent measurement of the muon $g-2$ and 
finalization of the LEP electroweak data, we re-examine all the 
new particle contributions to those observables in the MSSM. 
The SM fits the latest electroweak data excellently with the 
observed top-quark mass and the Higgs-boson mass of around 
$100\gev$, and so does the MSSM in the decoupling limit. 
The MSSM gives a slightly better fit to the data than the SM 
($\Delta \chi^2_{\rm min}\sim -2$) when relatively light 
left-handed sleptons of mass $\sim 200\gev$ and a light 
chargino of mixed higgsino-wino character $(\mu/M_2 \sim 1)$ 
with mass $\sim 100\gev$ co-exist. 
The improvement in the fit diminishes quickly for wino- or 
higgsino-dominant charginos, for heavier charginos, and for 
lighter sleptons. 
We find that the MSSM contributions to the muon $g-2$ is most 
efficient when the light chargino has a mixed character. 
If $\tan\beta < 10$, the set of a light mixed chargino and light 
sleptons that is favored by the electroweak data is also favored 
by the $g-2$ data for $\mu >0$. 
Models with light gauginos $(\mu/M_2 > 10)$ or light higgsinos 
$(\mu/M_2 < 0.1)$ give significant contributions to muon $g-2$ 
only for large $\tan\beta (\simgt 15)$. 
%%%%%----------------------------------  
\end{minipage}
\end{center}
\newpage
\baselineskip 18pt 
%%%%%%%%%%%%%%%%%%%%%%%%%%%%%%%%%%%%%%%%%%%%%%%%%%%%%%%%%%%%%%
%%%%%%%%%%%%         BEGINNING OF TEXT      %%%%%%%%%%%%%%%%%%
%%%%%%%%%%%%%%%%%%%%%%%%%%%%%%%%%%%%%%%%%%%%%%%%%%%%%%%%%%%%%% 
Looking for signatures of supersymmetry (SUSY) at high energy 
experiments is one of the most important tasks of particle physics. 
Since the LEP experiments have been completed without finding any 
evidences of new physics beyond the Standard Model (SM), the chance 
to discover the superparticles at the energy frontier has been 
postponed to the next stage, at the Tevatron Run-II or at the LHC. 
However, we may find important constraints on, or an indication of,  
the supersymmetric models from precise measurements of the SM 
particles which are sensitive to the interactions of superparticles. 
In this letter, we would like to examine the superparticle 
contributions to the muon $g-2$ in the light of its recent 
measurement~\cite{bnl01} and the finalization of the LEP 
electroweak data~\cite{lepewwg01}. 
The precise measurement of the muon $g-2$ has been achieved at 
BNL~\cite{bnl01} where the experimental uncertainty has been 
reduced by about factor 3 from the previous measurement~\cite{g-2old}.  
The current world average of the muon $g-2$ is then 
given by~\cite{bnl01}
%----
\bea
a_\mu({\rm exp.}) &=& 11~659~203(15) \times 10^{-10}. 
\eea
%----
The comparison of the data with the SM prediction is~\cite{bnl01}
%----
\bea
a_\mu({\rm exp.}) - a_\mu({\rm SM}) &=& 43(16) \times 10^{-10}, 
\label{eq:deviation}
\eea
%----
where the experimental and theoretical errors are added in 
quadrature. 
The theoretical uncertainty is dominated by the hadronic vacuum 
polarization effect. 
The SM prediction in eq.~(\ref{eq:deviation}) has been obtained 
by using the estimate of ref.~\cite{DH98}. 
Although consensus among experts should yet to emerge on the 
magnitude and the error of the SM prediction~\cite{yndurain}, 
we adopt the estimate of eq.~(\ref{eq:deviation}) as a distinct 
possibility. 
The purpose of this letter is to examine if this possible 
inconsistency of the data and the SM can be understood naturally 
in the context of minimal supersymmetric SM (MSSM), when taking 
account of the electroweak precision data.
%%%------------------------------

%%%------------------------------
The enormous data of the electroweak measurements at LEP1 
have been analyzed after its completion in 1995. 
The final combination of the results from four collaborations -- 
ALEPH, DELPHI, L3 and OPAL-- has been available on the 
$Z$-line shape and the leptonic asymmetry data~\cite{lineshape01}. 
The electroweak precision measurements consist of 17 $Z$-pole 
observables from LEP1 and SLC, and the $W$-boson mass from 
Tevatron and LEP2. 
%%%------------------------------

%%%------------------------------
Let us first summarize the constraints on the MSSM 
parameters~\cite{CH00} from the electroweak data. 
The supersymmetric particles affect these observables radiatively 
through the oblique corrections which are parametrized by 
$\sz, \tz, \mw$, and the $\zff$ vertex corrections 
$g_\lambda^f$, where $f$ stands for the quark/lepton species 
and $\lambda=L$ or $R$ stands for their chirality. 
The parameters $\sz$ and $\tz$~\cite{CH00} are related to 
the $S$- and $T$-parameters~\cite{stu90,HHKM} as:
%%%-----------------------------
\bsub
\bea
\dsz &=& \sz - 0.955 = \ds + \dr - 0.064 \xa 
         + 0.67 \frac{\ddelg}{\alpha}, 
\label{eq:dsz}
\\
\dtz &=& \tz - 2.65\hphantom{5} 
	= \dt + 1.49 \dr - \frac{\ddelg}{\alpha}, 
\label{eq:dtz}
\eea
\label{eq:sztzdr}
\esub
%%%-----------------------------
where $\dsz$ and $\dtz$ measure the shifts from the 
reference SM prediction point, 
$(\sz,\tz)=(0.955,2.65)$ at $\mt=175\gev,\mh=100\gev, 
\alps(\mz)=0.118$ and $1/\alpha(\mzsq)=128.90$. 
The $R$-parameter, which accounts for the difference between $T$ and 
$\tz$, represents the running effect of the $Z$-boson propagator 
corrections between $q^2=\mzsq$ and $q^2=0$~\cite{CH00}. 
The parameter $\xa \equiv \left( 1/\alpha(\mzsq)-128.90 
\right) /0.09$ allows us 
to take account of improvements in the hadronic uncertainty of 
the QED coupling $\alpha(\mzsq)$. 
$\ddelg$ denotes new physics contribution to the muon lifetime 
which has to be included in the oblique parameters because the Fermi 
coupling $G_F$ is used as an input in our formalism~\cite{CH00,HHKM}. 
The third oblique parameter $\dmw = \mw - 80.402({\rm GeV})$ is given 
as a function of $\ds,\dt,\du,\xa$ and $\ddelg$~\cite{CH00}. 
The explicit formulae of the oblique parameters and the vertex 
corrections $\Delta g_\lambda^f$ in the MSSM can be found 
in ref.~\cite{CH00}. 
%%%-------------

%%%------------- 
We study constraints on the oblique parameters from the 
electroweak data.  
In addition to the three oblique parameters, the $\zblbl$ vertex 
correction, $\dgbl$, is included as a free parameter 
in our fit 
because non-trivial top-quark-mass dependence appears only in 
the $\zblbl$ vertex among all the non-oblique radiative 
corrections in the SM. 
By using all the electroweak data~\cite{lepewwg01} and the 
constraint $\alps(\mz)=0.119\pm 0.002$~\cite{PDG00} on the QCD 
coupling constant, we find from a five-parameter fit 
($\dsz, \dtz, \dmw, \dgbl, \alps(\mz)$) 
the following constraints on the oblique parameters: 
%%%-------------
\bea
\begin{array}{l}
	\left. 
	\begin{array}{lcl}
	\Del S_Z -25.1 \Del g_L^b &=& \hph0.002 \pm 0.104 \\
\vsk{0.2}
	\Del T_Z -45.9 \Del g_L^b&=& -0.041 \pm 0.125
	\end{array} \right \}
\rho = 0.88, 
\\
\vsk{0.5}
\disp{ \Del m_W^{}({\rm GeV}) = \hph 0.032 \pm 0.037 },
%%%
%%%\\
%%%\vsk{0.5}
%%%\disp{
%%%\chi^2_{\sf min} = 22.6 + 
%%%\left(
%%%\begin{array}{c}
%%%\frac{\Del g_L^b + 0.00037}{0.00073}
%%%\end{array}
%%%\right)^2, 
%%%}
\end{array}
\eea
%%%------------- 
for $\dgbl = - 0.00037 \pm 0.00073$. 
The $\chi^2$ minimum of the fit is $\chi^2_{\rm min} = 22.6$ 
for the degree-of-freedom (d.o.f.) $19-5=14$. 
%%%------------- 

%%%------------- 
Through the expression (\ref{eq:dsz}) of $\dsz$, 
the QED coupling $\alpha(\mzsq)$ affects theoretical 
predictions for the electroweak observables. 
The LEP electroweak working group has adopted the new 
estimate~\cite{BP01} 
%-----
\bea 
1/\alpha(\mzsq)=128.936\pm 0.046~~(\xa = 0.4 \pm 0.51), 
\label{eq:alpha}
\eea
%-----
which takes into account the new $e^+ e^-$ annihilation 
results from BEPC~\cite{BES}. 
Using the central value of $\alpha(\mzsq)$ in eq.~(\ref{eq:alpha}), 
$\xa = 0.4$, the SM best fit is found given at 
$(\mt({\rm GeV}), \mh({\rm GeV}), \alps(\mz)) = (175.1, 116, 0.118)$. 
The $\chi^2$ minimum is $\chi^2_{\rm min}=24.4$ for 
the d.o.f. $20 - 3 = 17$. 
At the SM best fit point, the oblique parameters are given by 
$(\dsz-25.1\dgbl,\dtz-45.9\dgbl,\dmw) = (-0.010, -0.020, 
-0.009)$, which shows an excellent agreement with the data. 
Although the SM fit is already good, the further improvement 
of the fit may be found if new physics gives slightly positive 
$\dmw$. 
On the other hand, new physics contribution which gives 
large negative $\dsz$ and positive $\dtz$ is disfavored from 
the data. 
%----

%----
The supersymmetric contributions to the oblique parameters 
have been studied in ref.~\cite{CH00} in detail. 
In the MSSM, the oblique corrections are given as a sum of the 
individual contributions of (i) squarks, (ii) sleptons, 
(iii) Higgs bosons and the (iv) ino-particles (charginos 
and neutralinos). 
Squarks always give $\dsz \sim 0$ and $\dtz >0$ while sleptons 
give $\dsz \simlt 0$ and $\dtz >0$. 
Both of them give $\dmw>0$ which is favored from the data but 
the improvement is more than compensated by the disfavored 
contributions to $\dsz$ and $\dtz$. 
The contributions from the MSSM Higgs bosons are similar to 
that of the SM Higgs boson whose mass is around that of the 
lightest CP-even Higgs boson, as long as the CP-odd Higgs mass 
is not too small; $m_A \simgt 300\gev$~\cite{CH00}. 
We find no improvement of the fit through the oblique corrections 
in the Higgs sector. 
%%%-------

%%%------- 
The ino-particles give $\dtz <0$, owing to the large negative 
contribution 
to the $R$-parameter when there is a light chargino of mass 
$\sim 100\gev$~\cite{CH00}.  
They also make $\dsz$ negative when the light chargino is either 
gaugino-like or higgsino-like~\cite{CH00}. 
However, we find that both $\dsz$ and $\dtz$ can remain small 
in the presence of a light chargino, if the ratio of the 
higgsino mass $\mu$ 
and the SU(2)$_L$ gaugino mass $M_2$ is order unity~\cite{CH01}.  
Let us recall that $\sz$ is the sum of $S$- and $R$-parameters, 
while $\tz$ is the sum of $T$- and $R$-parameters. 
The $S$- and $T$-parameters are associated with 
the SU(2)$_L\times$U(1)$_Y$ 
gauge symmetry breaking while the $R$-parameter is negative as 
long as a light chargino of mass $\sim 100\gev$ exists. 
The contributions of the ino-particles to the $S$- and 
$T$-parameters are essentially zero when the lighter chargino 
is almost pure wino or pure higgsino, whereas 
they both become positive when their mixing is large 
because the mixing occurs through the gauge symmetry breaking. 
As a consequence, the negative $R$ contributions from a light 
chargino can be compensated by the positive $S$ and $T$ 
contributions to the parameters $\sz$ and $\tz$, if the 
light-chargino has a mixed charcter 
$(|\mu/M_2|\sim 1)$\footnote{All our numerical results are 
obtained under the constraint $M_2/\alpha_2 = M_1/\alpha_1$, 
although our results are insensitive to the magnitude of 
$M_1$.}. 
The parameter $\dmw$ is increased by the light ino-sector 
contribution when $\mu/M_2 \sim 1$, and hence the fit is 
slightly improved. 
This is largely because of the positive $T$ contribution 
due to the symmetry breaking. 
The overall fit, therefore, can be improved in the MSSM 
if a light chargino with the mixed wino-higgsino character 
$(|\mu/M_2 \sim 1|)$ exists and all sfermions are heavy. 
We find no sensitivity to the sign of the ratio $\mu/M_2$ 
in the fit to the electroweak data. 
%%%----------------------------------------  
%%%  		figure
%%%----------------------------------------    
\begin{figure}[t]
\begin{center}
%\rotatebox{90}{
\includegraphics[width=7cm,clip]{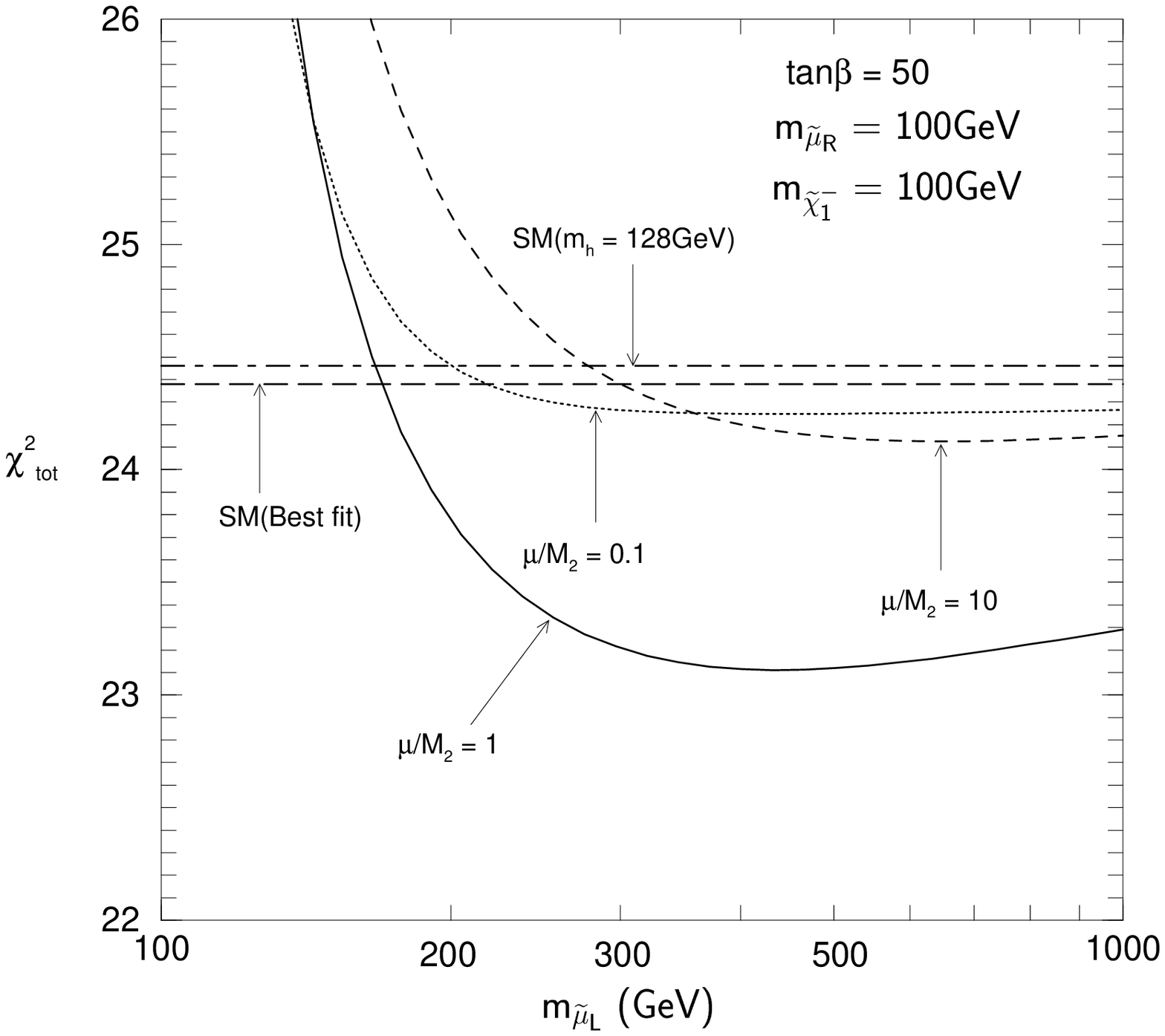}
\includegraphics[width=7cm,clip]{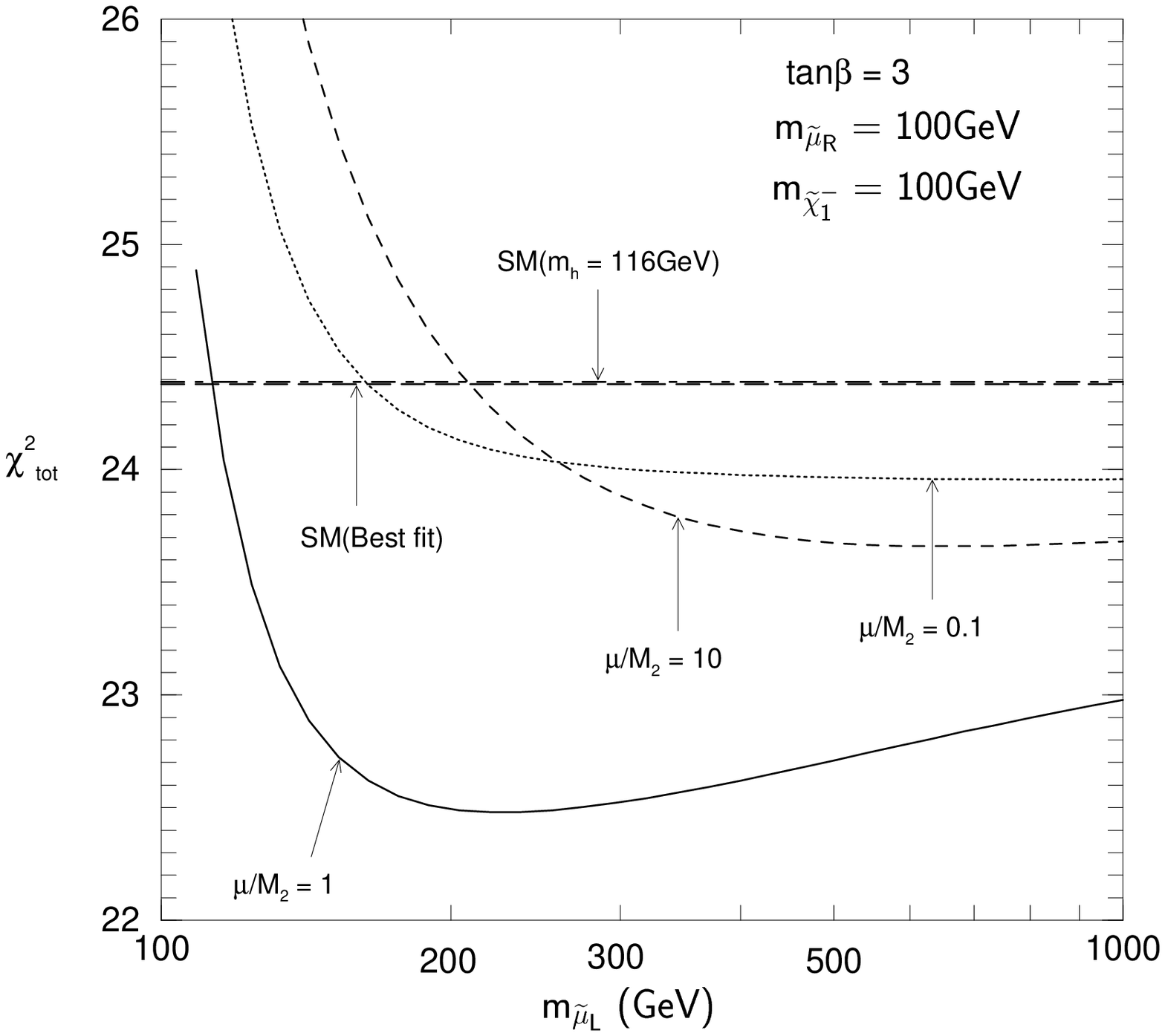}
%}
\caption{
Total $\chi^2$ in the MSSM as a function of the left-handed 
smuon mass $m_{\wt{\mu}_L}$ for $\tan\beta=50$ (a) and 
$3$ (b). 
Three lines are corresponding to $\mu/M_2 = 1.0$ (solid), 
0.1 (dotted) and 10 (dashed), respectively. 
The right-handed smuon mass $m_{\wt{\mu}_R}$ and the lighter 
chargino mass $m_{\wt{\chi}^-_1}$ are fixed at $100\gev$. 
The flavor universality of the slepton masses is assumed. 
The masses of squarks and extra Higgs bosons are set at $1\tev$. 
The SM best fit is shown by the dotted holizontal line. 
The dot-hashed horizontal line is the SM fit with $\mh=128\gev$ 
(left) and $\mh=116\gev$ (right)
which is the lightest Higgs mass predicted in the MSSM. 
}
\label{fig:global}
\end{center}
\end{figure}
%%%------------------

%%%------------------
Now we examine the effects of vertex and box corrections. 
Since we find that the light chargino can improve the fit 
slightly through the oblique corrections, we set the 
chargino mass to be $m_{\wt{\chi}^-_1} = 100\gev$, 
as a representative number in our analysis\footnote{
Our results are not significantly altered as long as the mass 
of the lighter chargino is smaller than about $150\gev$.}.  
Our task is to look for the possibility of further improving 
the fit through the vertex and box corrections when squarks 
and sleptons are also light. 
We find that sizable $\zff$ vertex corrections via the loop 
diagrams mediated by the left-handed squarks or the Higgs 
bosons make the fit worse always~\cite{CH01}. 
On the other hand, the fit is found to be improved slightly 
by the slepton contributions to the $Z\ell \ell$ vertices 
$(\Delta g_\lambda^\ell)$ and the muon lifetime $(\ddelg)$, 
when the left-handed slepton mass is around $200\sim 500\gev$. 
We show the total $\chi^2$ as a function of the left-handed 
smuon mass $m_{\wt{\mu}_L}$ in Fig.~\ref{fig:global}. 
The $\tan\beta$ dependence is shown for $\tan\beta=50$ (a) 
and $\tan\beta=3$ (b), and the character of the $100\gev$ 
lighter chargino is shown by $\mu/M_2=1.0$ (solid), 0.1 
(dotted) and 10 (dashed).  
For simplicity, we assume that the universality of the 
slepton mass parameters in the flavor space. 
We find no sensitivity to the right-handed slepton mass, 
and $m_{\wt{\mu}_R}$ is fixed at $100\gev$. 
The masses of all the squarks and the CP-odd Higgs-boson 
mass are set at $1\tev$. 
The improvement of the fit is maximum at around $m_{\wt{\mu}_L} 
\simeq 300\gev$ for $\tan\beta=50$ (a) and $\simeq 200\gev$ 
for $\tan\beta=3$ (b) for $\mu/M_2=1.0$, where the total 
$\chi^2$ value is smaller than those of the decoupling 
limits, which are shown by the dot-dashed horizontal lines, 
by about 1.4 and 1.9, respectively. 
%%%----

%%%----
The origin of the improvement at those points is found to 
come from the vertex corrections to the hadronic peak cross 
section on the $Z$-pole, which more than compensate the 
disfavored negative contributions to the oblique parameter 
$\sz$ from the light left-handed sleptons~\cite{CH00}. 
The hadronic peak cross section $\sigmah$ is given by 
%----
\bea
\sigmah &=& \frac{12\pi}{\mzsq} 
         \frac{\Gamma_e\Gamma_h}{\Gamma_Z^2}, 
\eea
%---- 
and is almost independent of the oblique 
corrections~\cite{CH00, Hagiwara}. 
The final LEP1 data of $\sigmah$ is larger than the SM 
best-fit value by about 2-$\sigma$.  
Since the squarks and Higgs bosons are taken to be heavy, 
the partial decay width $\Gamma_e$ is the only quantity which 
is affected significantly by the vertex corrections. 
The supersymmetric contribution to $\Gamma_e$ is given by 
the sleptons and the ino-particles, which constructively 
interferes with the SM prediction. 
The fit to the $\sigmah$ data, therefore, improves if the 
sleptons and ino-particles are both light. 
The overall fit is found to improve when the left-handed slepton 
mass is around $200\sim 500\gev$, as shown in Fig.~\ref{fig:global}. 
If the slepton mass is too light ($m_{\wt{\mu}_L} \simlt 200\gev$ 
for $\tan\beta=50$, $m_{\wt{\mu}_L} < 150\gev$ for $\tan\beta=3$), 
the total $\chi^2$ increases rapidly because of disfavored 
contributions to the $\sz$-parameter and also from the muon 
lifetime $(\ddelg)$. 
Since, in the large $m_{\wt{\mu}_L}$ limit, only the oblique 
corrections from the light chargino and neutralinos remain, 
the difference of $\chi^2$ between the value at its minimum 
and that at $m_{\wt{\mu}_L} \sim 3\tev$ represents the improvement 
of the fit due to non-oblique corrections. 
Among the three cases of $\mu/M_2$ in Fig.~\ref{fig:global}, 
only the $\mu/M_2=1.0$ case shows a slight improvement of the 
fit via the non-oblique corrections. 
This is because the relatively light heavier chargino 
($m_{\wt{\chi}^-_2}\simeq 220\gev$ for $\mu/M_2=1$) 
contributes to $\Gamma_e$ but has no other significant effects 
elsewhere. 
The improvement of the fit persists for $\mu/M_2 \sim 0.5$ or 2, 
but the smallest $\chi^2$ is found at $\mu/M_2=1.0$. 
Although we have shown results for $\mu/M_2>0$, we found that 
the electroweak data are insensitive to the sign of $\mu/M_2$. 
%%%--------------- 
\begin{figure}[t]
\begin{center}
\includegraphics[width=10cm,clip]{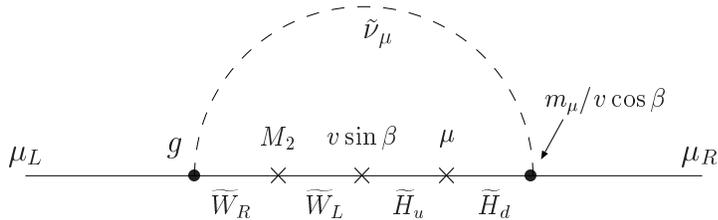}
\caption{
Feynman diagram for the muon $g-2$ which is mediated by a 
sneutrino and charginos. 
The photon line should be attached to the charged particle 
lines. 
}
\label{fig:diagram}
\end{center}
\end{figure}
%%%------------------

%%%---------------
Now we are ready for the study of the supersymmetric contribution 
to the muon $g-2$ in the light of the electroweak precision data. 
The above study tells us that the electroweak precision data 
favors the presence of relatively light charginos 
$(m_{\wt{\chi}^-_1} \sim 100\gev)$ of mixed wino-higgsino 
character $(|\mu/M_2| \sim 1)$, and that the co-existence 
of left-handed sleptons can further improve the fit 
when its mass is in the $200\sim 500\gev$ range. 
This is in fact the region of the MSSM parameter space 
which gives a sizable contributions to the muon 
$g-2$~\cite{g-2susy:one,CHH,g-2susy:two}. 
It is known that the MSSM contribution to $a_\mu$ is most 
efficient when $|\mu/M_2| \sim 1$~\cite{g-2susy:one,CHH}, 
and the sign of 
the discrepancy between the data and the SM prediction 
in eq.~(\ref{eq:deviation}) favors positive $\mu/M_2$. 
This may be understood intuitively from the diagram of 
Fig.~\ref{fig:diagram}, where the $\mu_L$-$\mu_R$ transition 
amplitude is expressed in terms of the electroweak symmetry 
eigenstates. 
Since the muon $g-2$ is given as the coefficient of the magnetic 
dipole operator, the chirality of the external muon must be 
flipped at somewhere. 
In the chargino-sneutrino exchanging diagram, the chirality flip 
occurs at the internal fermion line. 
We can tell from the diagram of Fig.~\ref{fig:diagram} that 
the relevant MSSM contribution to the muon $g-2$ is proportional 
to the product $M_2 \mu \tan\beta$. 
In the wino or higgsino limit, the contribution is suppressed 
because one of the two charginos is heavy. 
The chirality flip due to $\wt{\mu}_L$-$\wt{\mu}_R$ mixing 
contributes negligibly to $a_\mu$ even at 
$\tan\beta=50$~\cite{CHH}. 
In the following analysis, we therefore ignore the small 
$\wt{\mu}_L$-$\wt{\mu}_R$ mixing effects. 
%%%------------------
\begin{figure}[t]
\begin{center}
\includegraphics[width=16cm,clip]{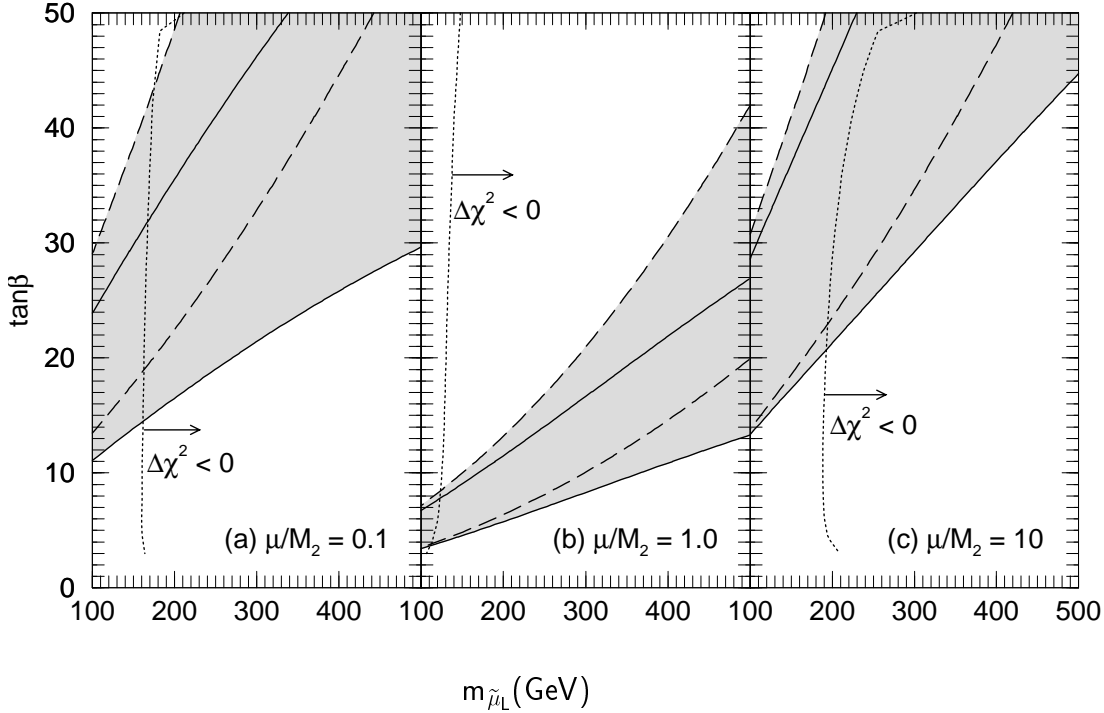}
\caption{
1-$\sigma$ allowed region of the $(m_{\wt{\mu}_L}, \tan\beta)$ 
plane from the experimental data of muon $g-2$ (\ref{eq:deviation}) 
for $\mu/M_2 = 0.1$~(a), 1.0~(b) and 10~(c). 
The lighter chargino mass $m_{\wt{\chi}^-_1}$ is set at $100\gev$. 
The region enclosed by the solid lines and dashed lines are 
obtained for the right-handed smuon mass $m_{\wt{\mu}_R}=100\gev$ 
and $500\gev$, respectively. 
In the region of $m_{\wt{\mu}_L}$ smaller than the vertical line, 
the MSSM fit to the electroweak data is worse than the SM 
$(\Del\chi^2 \equiv \chi^2_{\rm min}\left[{\rm MSSM}\right]- 
\chi^2_{\rm min}\left[{\rm SM}\right] > 0)$. 
}
\label{fig:three}
\end{center}
\end{figure}
%%%------------------

%%%------------------
In Fig.~\ref{fig:three}, we show constraints on the left-handed 
smuon mass $m_{\wt{\mu}_L}$ and $\tan\beta$ from the experimental 
data of $a_\mu$, eq.~(\ref{eq:deviation}), for 
$m_{\wt{\chi}^-_1}=100\gev$ 
and $\mu/M_2=0.1$ (a), 1.0~(b) and 10~(c). 
The regions enclosed by solid and dashed lines are found for the 
right-handed smuon mass $m_{\wt{\mu}_R}=100\gev$ and $500\gev$, 
respectively. 
In the region of $m_{\wt{\mu}_L}$ smaller than the vertical 
dotted lines, the MSSM fit to the electroweak data is worse than 
the SM fit 
($\chi^2_{\rm min}\left[{\rm MSSM}\right] > 
\chi^2_{\rm min}\left[{\rm SM}\right]$).  
Fig.~\ref{fig:three} (b) tells us that if the lighter chargino 
state has comparable amounts of the wino and higgsino components 
$(\mu/M_2=1.0)$, which is favored from the electroweak data, 
relatively low values of $\tan\beta$ is allowed: 
$4 \simlt \tan\beta \simlt 8$ for $m_{\wt{\mu}_L}\approx 110\gev$. 
This is the region favored by the electroweak data in Fig.~\ref{fig:global}. 
On the other hand, if it is mainly higgsino $(\mu/M_2 = 0.1)$ or 
wino ($\mu/M_2=10$), low values of $\tan\beta$ is excluded: 
$\tan\beta \simgt 15$ for $m_{\wt{\mu}_L} \approx 200\gev$, 
where the MSSM fit to the electroweak data is comparable to 
the SM. 
In all cases of $\mu/M_2$ in the figure, the right-handed smuon 
tends to make the bound on $\tan\beta$ lower if its mass is small. 
It should be noted that the current $g-2$ data can be explained 
even for larger $m_{\wt{\mu}_L}$ for appropriately large 
$\tan\beta$. 
The electroweak data is insensitive to $m_{\wt{\mu}_L}$ in 
this region. 
%%%------------------

\begin{figure}[t]
\begin{center}
\includegraphics[height=7cm,clip]{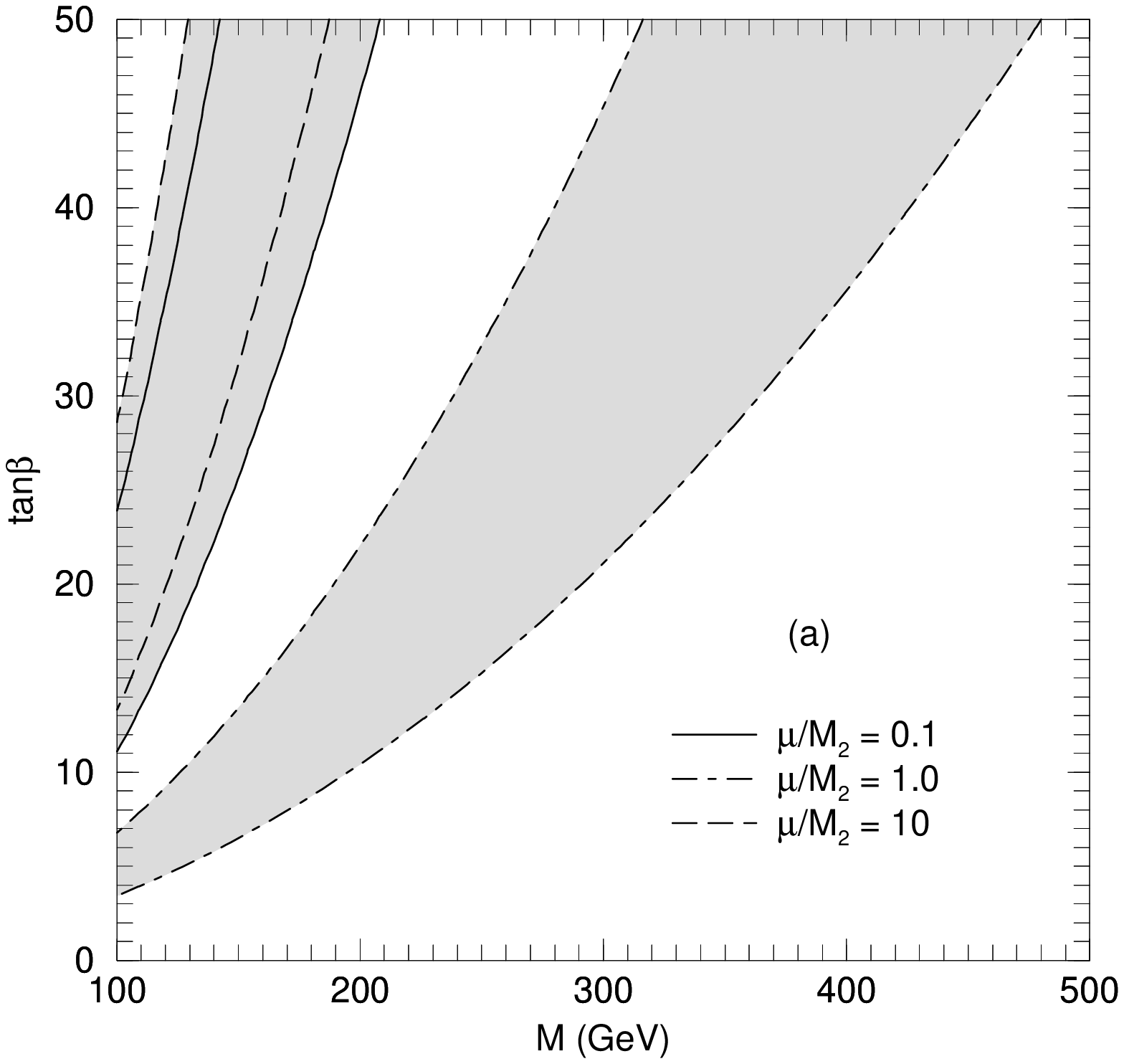}
\includegraphics[height=7cm,clip]{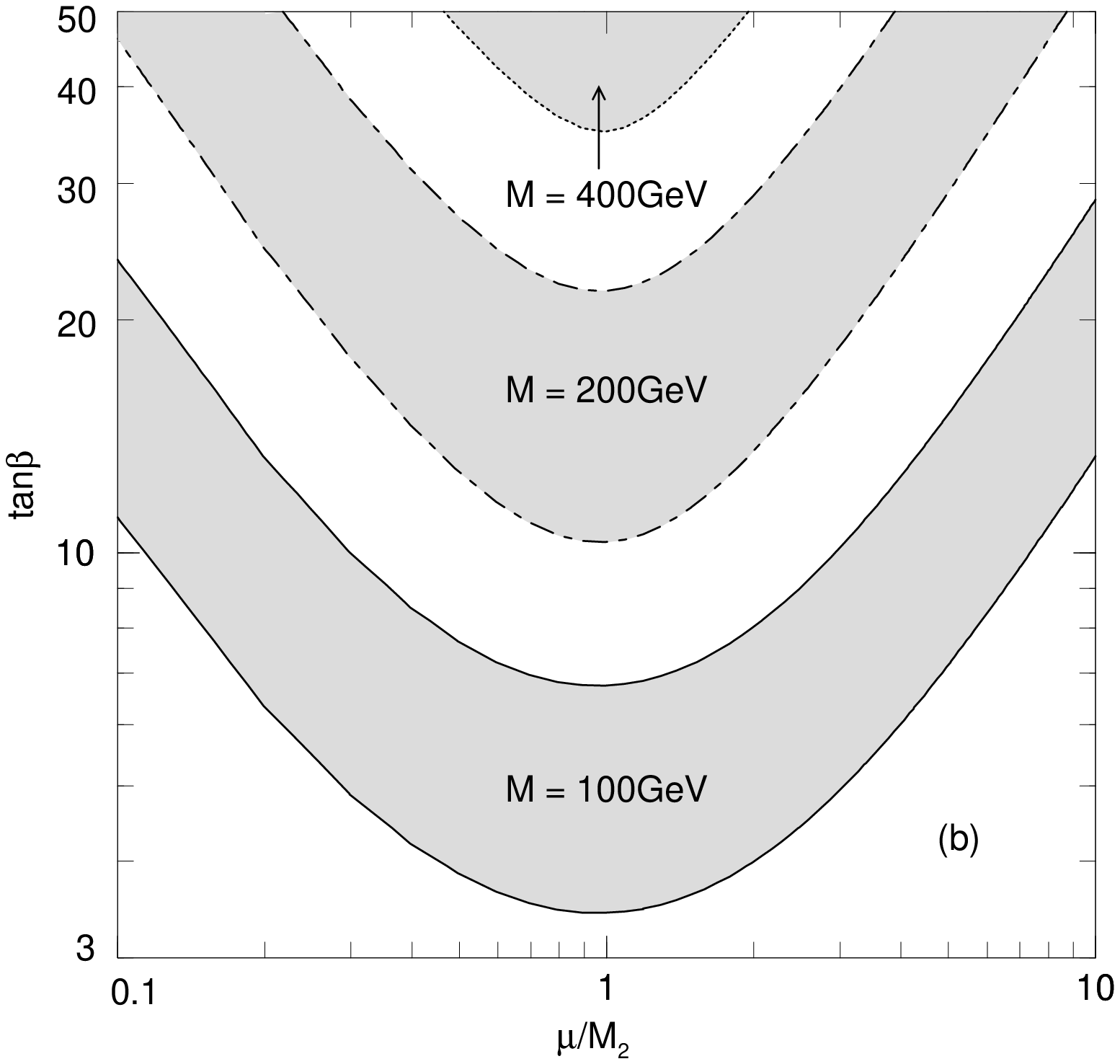}
\caption{
Constraints from the muon $g-2$ data (\ref{eq:deviation}) 
on ($M, \tan\beta$) (a) 
and on ($\mu/M_2, \tan\beta$) (b).  
The mass parameter $M$ is defined as 
$M \equiv m_{\wt{\chi}^-_1}= m_{\wt{\mu}_L}= m_{\wt{\mu}_R}$. 
The solid, dot-dashed and dashed lines are corresponding 
to: (a) $\mu/M_2 = 0.1$, 1.0 and 10, and 
(b) $M=100\gev$, $200\gev$ and $400\gev$, respectively. 
}
\label{fig:four}
\end{center}
\end{figure}

%%%------------------
In Fig.~\ref{fig:four}, we show the 1-$\sigma$ allowed range 
from the muon $g-2$ data (\ref{eq:deviation}) when all the 
relevant charged superparticles have the common mass, 
$m_{\wt{\chi}^-_1}= m_{\wt{\mu}_L}= m_{\wt{\mu}_R} \equiv M$. 
The constraints on $(M,\tan\beta)$ are given in 
Fig.~\ref{fig:four}(a), for three representative cases of 
$\mu/M_2=0.1,1.0$ and 10. 
Because the muon $g-2$ decreases if any of the three charged 
superparticles is heavier than the common value $M$, we can 
regard the allowed range as an upper mass limit of charged 
superparticles. 
We find that, if the ligther chargino is either 
wino- or higgsino-like, \ie, $\mu/M_2 \gg 1$ or $\mu/M_2 \ll 1$, 
either the lighter chargino or smuons should be discovered by 
a lepton collider at $\sqrt{s}=400\gev$ for any $\tan\beta (\le 50)$. 
On the other hand, if no superparticle is found at a $500\gev$ 
lepton collider, then the chargino should have the mixed 
character and $\tan\beta$ should be bigger than 15. 
%%%------------------
%%%------------------
In Fig.~\ref{fig:four}(b) we show the constraint on 
$(\mu/M_2,\tan\beta)$ from the $a_\mu$ data (\ref{eq:deviation}). 
We can clearly see from this figure that the lowest value of 
$\tan\beta$ is allowed at $\mu/M_2 = 1$. 
%%%------------------ 

%%%------------------
To summarize, we have studied the supersymmetric contributions 
to the muon $g-2$ in the light of its recent measurement 
at BNL and the finalization of the LEP electroweak data. 
Although the SM fit to the electroweak data is good, 
slightly better fit in the MSSM is found when relatively 
light left-handed sleptons with mass $\sim 200\gev$ and a light 
chargino of mass $\sim 100\gev$ and of mixed wino-higgsino character 
($\mu/M_2 \sim 1$) exist. 
The improvement is achieved via the light chargino contribution 
to the oblique parameters and also via the ino-slepton contribution 
to $\sigmah$. 
The improvement of the fit disappears rapidly if the light 
chargino is higgsino- or wino-like, or if the light chargino mass 
is heavier $(\simgt 200\gev)$, or the sleptons are too light 
($\simlt 180\gev$ for $\tan\beta=50$ and 
$\simlt 120\gev$ for $\tan\beta=3$). 
We find that the supersymmetric contribution to the muon $g-2$ 
is most efficient for $\mu/M_2 \sim 1$. 
If $\tan\beta \simlt 10$, the MSSM parameter space which is favored 
from the electroweak data is also favored from the muon $g-2$ data. 
The wino- or higgsino-dominant chargino contributes significantly 
to the muon $g-2$ only for large $\tan\beta$ ($\simgt 15$), 
although it does not improve the fit to the electroweak data. 
The impact on the search for the superparticles at future 
colliders from the precise measurement of the muon $g-2$ is 
also discussed. 
The present 1-$\sigma$ constraint (\ref{eq:deviation}) 
from the muon $g-2$ measurement implies that either a chargino 
or charged sleptons are within the discovery limit of a 
$500\gev$ lepton collider for any $\tan\beta (< 50)$ 
if the lighter chargino is dominantly wino $(\mu/M_2 \simgt 3)$ 
or dominantly higgsino $(0< \mu/M_2 \simlt 0.3)$. 
\vspace{0.5cm}

{\bf \Large Acknowledgements}
\vspace{0.3cm}

The authors thank S. Eidelman, M. Hayakawa and J.H. K\"{u}hn 
for fruitful discussions. 
%-------------------------------------

%%%%%%%%%%%%%%%%%%%%%%%%%%%%%%%%%%%%%%%%%%%%%%%%%%%%%%%%%%%%%%
%%%%%%%         BIBLIOGRAPHY           %%%%%%%%%%%%%%%%%%%%%%%
%%%%%%%%%%%%%%%%%%%%%%%%%%%%%%%%%%%%%%%%%%%%%%%%%%%%%%%%%%%%%% 

%
\end{document}